\documentclass[useAMS,usenatbib]{mn2e}
\usepackage{epsfig}

\title[Critical Mass Transfer in Double-Degenerate Type~Ia
Supernovae]{Critical Mass Transfer in Double-Degenerate Type~Ia Supernovae}

\author[R. G. Martin, C. A. Tout and P. Lesaffre]{Rebecca
G. Martin$^1$\thanks{E-mail:
rgm32@cam.ac.uk; cat@ast.cam.ac.uk; lesaffre@ast.cam.ac.uk},
Christopher A. Tout$^{1,2}$ and Pierre Lesaffre$^1$\\
$^1$University of Cambridge, Institute of Astronomy, The Observatories,
Madingley Road, Cambridge CB3 0HA\\
$^2$Centre for Planetary and Stellar Astrophysics, School of
Mathematics, Monash University, Clayton, Victoria 3800, Australia}
\begin{document}

\date{}

\pagerange{\pageref{firstpage}--\pageref{lastpage}} \pubyear{2005}
\maketitle

\label{firstpage}

\begin{abstract}
Doubly-degenerate binary systems consisting of two white dwarfs both
composed of carbon and oxygen and close enough that mass is
transferred from the less massive to the more massive are possible
progenitors of type~Ia supernovae.  If the mass transfer rate is slow
enough that the accreting white dwarf can reach a mass of $1.38\,\rm
M_{\odot}$ then it can ignite carbon degenerately at its centre.  This
can lead to a thermonuclear runaway and thence a supernova explosion.
However if the accretion rate is too high the outer layers of the
white dwarf heat up too much and carbon ignites there
non-degenerately.  A series of mild carbon flashes can then propagate
inwards and convert the carbon to neon relatively gently.  There is no
thermonuclear runaway and no supernova.  We examine the critical
accretion rate at which ignition switches from the centre to the
surface for a variety of white dwarfs and find it to be about two
fifths of the Eddington rate.  In a real binary star the mass transfer
rate falls off as mass transfer proceeds and the system widens.  Even
if the initial transfer rate is high enough for carbon to ignite at
the outer edge, if such rapid accretion were to persist, we find that
it can extinguish if the rate drops sufficiently quickly.  The
interior of the white dwarf remains carbon rich and, if sufficient
mass can still be transferred from the companion, it can eventually
ignite degenerately at the centre.  The primary white dwarf must be
about $1.1\,M_\odot$ or above and the companion about $0.3\,M_\odot$.
Though white dwarfs of such low mass are expected to be pure helium we
note that a star of initial mass $2.5\,M_\odot$ has a CO core of about
$0.3\,M_\odot$ when it begins to ascend the asymptotic giant branch.
Alternatively if the accretion rate can be limited to a maximum of
$0.46$ of the Eddington rate then a $1.1\,M_\odot$ white dwarf
accretes sufficiently slowly to explode from a companion white dwarf
of any large enough mass.

\end{abstract}


\section{Introduction}
Type~Ia supernovae (SNe~Ia) are the brightest objects in normal
galaxies. They appear to be standard candles and hence are useful
cosmological measures of the Universe \citep{Baa}. They are also a
major source of iron \citep{tout2001} which tends to be trapped in the
neutron star remnants of other supernovae.  However, their progenitors
remain uncertain (see for example \citet{tout2005}) and this leaves
their standard nature questionable particularly in the light of the
chemical evolution of the Universe. 

Type~I supernovae have no hydrogen lines and type~Ia are further
distinguished by their prominent silicon lines.  They are almost
certainly exploding carbon and oxygen (CO) white dwarfs.  Their available
nuclear energy exceeds their binding energy so that the
whole star can be destroyed in a thermonuclear runaway.  Most of the
material approaches nuclear statistical equilibrium and about
$0.6\,\rm M_{\odot}$ of $^{56}\rm Ni$ is expelled \citep{ropke2005}.
The radioactive decay of $^{56}$Ni to $^{56}$Fe via $^{56}$Co powers
the supernova and the decay signature has been clearly observed
\citep{branch1998}.  It is fairly certain that SNe~Ia are detonated by
mass accretion on to the CO white dwarf.  As the mass of a cold white
dwarf increases towards the Chandrasekhar mass, the gravitational
collapse heats the material and fusion ignites in the degenerate core.
Typically CO white dwarfs explode at about $1.38\,
\rm M_{\odot}$.

Accreting white dwarfs have been known for some time as the engines of
cataclysmic variables, the source of novae and dwarf novae
\citep{warner1995}, and so were the first candidates to be considered.
However if the accreting material is hydrogen-rich, accumulation of a
layer of only $10^{-5}-10^{-3}\,M_\odot$ of cold material leads to
degenerate ignition of hydrogen burning sufficiently violent to eject
most, if not all of or more than, the accreted layer in the well known
nova outbursts of cataclysmic variables.  The white dwarf mass does
not significantly increase and ignition of its interior is usually
avoided.  However if the accretion rate is high, $\dot M >
10^{-7}\,M_\odot\,\rm yr^{-1}$, compressional heating of the surface
layers raises the degeneracy and hydrogen can burn relatively gently as
it is accreted, bypassing novae explosions
\citep{paczynski1978}, allowing the white dwarf mass to
grow.  Though, if it is not much larger than this, $\dot M > 3\times
10^{-7}\,M_\odot\,\rm yr^{-1}$, hydrogen cannot burn fast enough so
that accreted material builds up a giant-like envelope around the core
and burning shell which eventually leads to more drastic interaction
with the companion and probably the end of the mass transfer episode.
Rates in the narrow range for steady burning are found only when the
companion is in the short-lived phase of thermal-timescale expansion
as it evolves from the end of the main sequence to the base of the
giant branch.  Super-soft X-ray sources \citep{kahabka1997} are
probably in such a state but cannot be expected to remain in it for
very long (a SNe~Ia rate of only about one millionth of the observed
rate form by this channel in the population synthesis of
\cite{hurley2002}) and white dwarf masses almost never increase
sufficiently to explode as SNe~Ia.  If He-rich material is accreted
instead about $0.1 - 0.15\,M_\odot$ of degenerate material can
accumulate before ignition
\citep{nomoto1982} and an accretion rate above $3\times
10^{-8}\,M_\odot\,\rm yr^{-1}$ can lead to steady burning
\citep{saio1987}.
\par
The currently popular model overcomes these
problems by postulating that, when the mass-transfer rate exceeds that
allowed for steady burning, only just the right fraction of the mass
transferred is actually accreted by the white dwarf.  A viable
mechanism for this is a strong wind from the accretion disc that
expels material from the system before it reaches the white dwarf
\citep{hachisu1996}.  Alternatively the white dwarf might indeed swell
up to giant dimensions but the resulting common-envelope evolution
could be sufficiently efficient that the small amount of excess
material can be ejected without the cores spiralling in.  This is
quite consistent with the findings of \citet{nelemans2005} and
\citet{nelemans2000} that such efficiency
necessary for at least one phase of common-envelope evolution in the
formation of close double white dwarf systems.  Helium-accreting CO
white dwarfs were
once popular when it was thought that a thermonuclear runaway in the
accreted helium layer could set off the CO core in an edge-lit
detonation \citep{woosley1994}.  However 2--D models indicate that
central ignition is unlikely and light curves and spectra do not fit well
so they are no longer considered viable
progenitors \citep{branch1995}.

In this work we consider the alternative that many complications can
be avoided altogether if the white dwarf can accrete material of
similar composition to itself,
carbon and oxygen.  This can
be achieved in a double degenerate model in which two CO white dwarfs
merge \citep{Webbink1984}.  Both form as the cores of asymptotic giant
branch stars and, after one or two phases of mass transfer, one or
both of which involves substantial orbital shrinkage through
common-envelope evolution, they are brought close enough together that
angular momentum losses owing to gravitational radiation lead to Roche
lobe overflow.  A major difficulty with this model lies in the fact
that high accretion rates also heat the white dwarf.  This heating is
due mainly to the gravitational compression of the white-dwarf
material as the mass increases.  The actual accretion luminosity
liberated by the material falling down the potential well on to the
white-dwarf surface has a negligible heating effect in comparison and
is in any case mostly liberated in an accretion disc or boundary layer
and radiated away.  A white dwarf can lose heat by conduction to the surface and
radiation or by interior neutrino loss processes such as
electron-positron pair production and annihilation. The material at
the surface undergoes the most rapid compression as it becomes
degenerate. At slow rates energy released in these surface layers can
escape fast enough but at high rates the surface layers reach
temperatures required for fusion to begin. True ignition occurs when
energy production by carbon fusion exceeds that lost in neutrinos, a
situation that is rapidly followed by the onset of convection but at
velocities that are not sufficient to carry away igniting packets
before thermonuclear runaway ensues. \cite{nomoto1985} calculated that
carbon ignites near the surface at constant accretion rates in excess
of about one fifth of the Eddington rate (see section~\ref{secedd}).
Ignition at the surface raises the degeneracy so that burning of
carbon to neon and sodium proceeds relatively gently.  Surface burning
triggers the gentle ignition of a deeper shell and carbon is thus
burnt successively throughout the white dwarf. Once the central carbon
has been exhausted the core can't ignite until hot enough for neon
burning by photodisintegration or oxygen fusion but by this point
electron capture by magnesium has taken the collapse beyond the point
at which the available nuclear energy can explode the white dwarf. It
simply collapses quietly to a neutron star releasing energy in
neutrinos. Thus for a SN~Ia a thermonuclear runaway must begin in
the core before carbon ignites sufficiently at the surface.

Because white dwarfs expand as they lose mass, unless the Roche lobe of
the mass donating companion expands even faster the very process of
mass transfer causes the mass-losing (lower mass) white dwarf to
overfill its Roche lobe yet more and mass transfer accelerates to
dynamical rates.  Only when the
mass ratio $q=M_{\rm loser}/M_{\rm accretor}<0.628$ can this positive
feedback be avoided.  It is not known what the final outcome of
dynamical mass transfer is but it is likely that much of the loser is
lost to the interstellar medium and that the remainder is accreted at
very high rates.  This already excludes the majority (about
four-fifths in the population synthesis models of \cite{hurley2002}) of merging
CO white dwarf pairs which tend to have component masses between $0.5$
and~$1.1\rm M_{\odot}$. In the remaining cases, where dynamical mass transfer
is avoided, the initial rate is still in excess of \cite{nomoto1985}'s limit
and so merging CO white dwarfs are not currently favoured
SNe~Ia progenitors.

Various mechanisms might force merging white dwarfs to spin rapidly.
\cite{piersanti2003a,piersanti2003b} modelled the effects of rapid
spin on carbon ignition in CO white dwarfs.  They found that
compressional heating was not much affected but that the rate of
diffusion of heat to the centre was slowed.  This generally reduces
the maximum rate of accretion that allows central carbon ignition.  On
the other hand spinning near breakup prevents accretion but also
distorts the white dwarf so that it spins down by its own
gravitational wave emission, while accreting just enough mass to keep
it spinning close to break up.  This leads to a natural limit to the
accretion rate of about $4\times 10^{-7}\,M_\odot\,\rm yr^{-1}$, low
enough to allow central carbon ignition.  In this work we only
consider slowly rotating white dwarfs because we are interested in
dynamically stable mass transfer.  For instance, when a $0.3\,M_\odot$
white dwarf begins mass transfer to a $1.1\,M_\odot$ white dwarf (as
in section~\ref{secsup}) the period of the binary is $146\,$s, well
above the break-up spin of the $1.1\,M_\odot$ white dwarf of $5.6\,$s.
It might however be argued that accretion from the inner edge of a
Keplerian accretion disc would still spin up the massive white dwarf
before it reached its ignition mass.  However this is avoided if the
white dwarf has a magnetic field strong enough to disrupt the
accretion disc as in the polar and intermediate-polar cataclysmic
variables \citep{warner1995}.  Indeed more massive white dwarfs are
thought to form with stronger magnetic fields
\citep{wickramasinghe2005}.  We thus expect the accreting white dwarfs
which interest us here to be relatively slowly rotating and so can be
modelled without rotation.

Here we reconsider the limit on the accretion rate.  After describing
the details of our evolution code and how we identify carbon ignition
(section~\ref{seccode}) we first work with constant rates
(section~\ref{seccon}) for direct comparison with \cite{nomoto1985}.
With modern computer power we can calculate on a much finer grid of
initial masses and accretion rates.  Secondly, we consider accretion
at a constant fraction of the Eddington rate (section~\ref{secedd})
and find somewhat higher limits than \cite{nomoto1985}.  Finally we consider
true binary evolution in which the rate is initially high but falls
off as the donor white dwarf loses mass and the system widens
(section~\ref{secbin}).  In section~\ref{secsup} we explore the range
of initial binary star parameters for which we find central carbon
ignition and thence a supernovae and in section~\ref{secconc} we
present our conclusions and consider whether realistic progenitor
systems might exist.

\section{Stellar Evolution Models}
\label{seccode}

We use the Cambridge STARS code to make, evolve and accrete on to
white dwarfs.  The STARS code is the most recent version
of the Eggleton evolution program \citep{Egg71,Egg72,Egg73}. The
equation of state, which includes molecular hydrogen, pressure
ionization and Coulomb interactions is discussed by
\cite{Pols}.  We take the initial composition of the star to be uniform
with a hydrogen abundance $X =0.7$, helium $Y=0.28$ and metals
$Z=0.02$. The metal mixture is according to the meteoritic abundances determined
by \cite{ander}.  Only nuclear burning that affects the structure is
included, hydrogen burning by the p-p chain and the CNO cycles, helium
burning by the triple-$\alpha$ reaction and reactions with $\rm
^{12}C$, $\rm ^{14}N$ and $\rm ^{16}O$ and carbon burning via $\rm
^{12}C+{^{12}C}$ only.  Other isotopes and reactions are not explicitly
modelled. The reaction rates are taken from \cite{cau} and opacity
tables are from \cite{Igles} and \cite{alex}. An Eddington
approximation \citep{Wool} is used for the surface boundary conditions
at an optical depth of $\tau =2/3$.

We initially construct white dwarf models from those of asymptotic
giant branch stars with appropriate core masses by stripping them of
their envelopes at very high mass-loss rates to mimic the binary
formation process.  This leaves CO cores which can cool, without
further mass loss, to become white dwarfs.  We obtain white dwarfs of
different temperatures by varying the cooling time and we can form
white dwarfs with different compositions by evolving to the AGB from
main-sequence stars of different initial mass (see tables \ref{tab1}
and~\ref{eddtab}).

Once we have a white dwarf we can accrete mass back on to it. However
the mass we accrete now is of the same composition as the surface of
the white dwarf.  There are then no problems with numerical diffusive
mixing.  Because our evolution code is hydrostatic, to ensure
numerical convergence, we must prime our white dwarfs for rapid
accretion by increasing the
rate from zero to the desired amount in several short steps.
These steps are short enough that this is effectively instantaneously compared with the thermal
timescale in the regions of interest.

\begin{figure}
\epsfxsize=8.4cm
\epsfbox{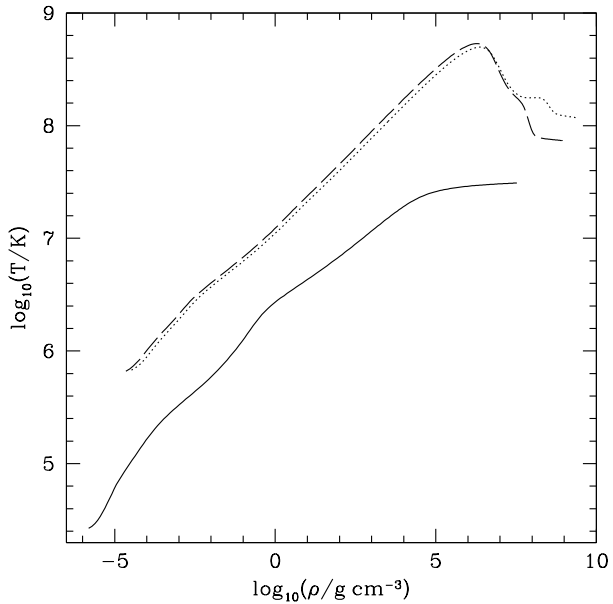}
\caption[]{The internal temperature density profile of a white dwarf
of initial mass $1.2\,\rm M_{\odot}$ with a central temperature of
$10^{7.5}\,\rm K$ made from the core of a $5\rm M_{\odot}$ star.  The
solid line is this initial state before accretion.  Nearly all the
mass lies in the isothermal core.  Most of the variation in
temperature occurs through the very thin atmosphere.  The dotted line
has accreted to $1.38\,M_\odot$ at $2\times 10^{-6}\,M_\odot\,\rm
yr^{-1}$ and is about to ignite carbon at its centre.  The dashed line
has accreted to $1.35\,M_\odot$ at $3\times 10^{-6}\,M_\odot\,\rm
yr^{-1}$ and is about to ignite carbon just below its surface.  The
ignition of each of these is shown in more detail in figures \ref{ign}
and~\ref{ign2}.}
\label{initial}
\end{figure}

\begin{figure}
\epsfxsize=8.4cm
\epsfbox{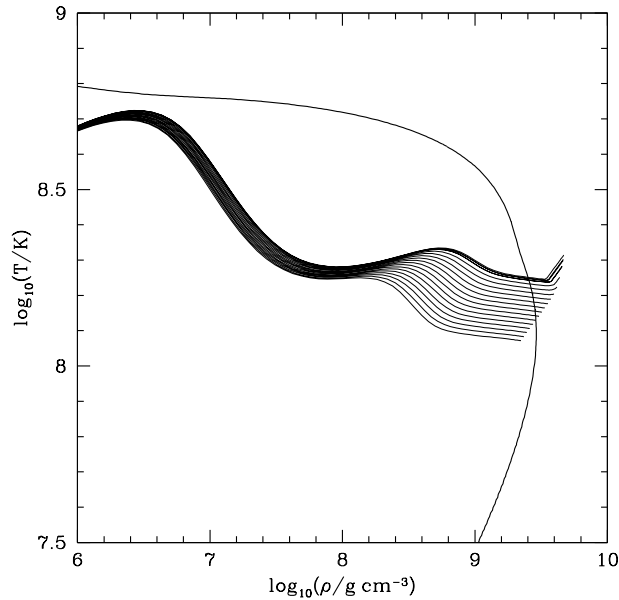}
\caption[]{The line crossing both axes in the carbon ignition curve on which
energy generation by carbon burning equals the energy loss in
neutrinos.  The other lines are the run of temperature and density
through the white dwarf, at late times, increasingly later from bottom
to top, beginning with the dotted line of figure~\ref{initial}.  In this case these cross the ignition curve at the centre and
a thermonuclear runaway can begin there under degenerate conditions.
The accretion rate is $2\times 10^{-6}\rm \, M_{\odot}yr^{-1}$.}
\label{ign}
\end{figure}

\begin{figure}
\epsfxsize=8.4cm
\epsfbox{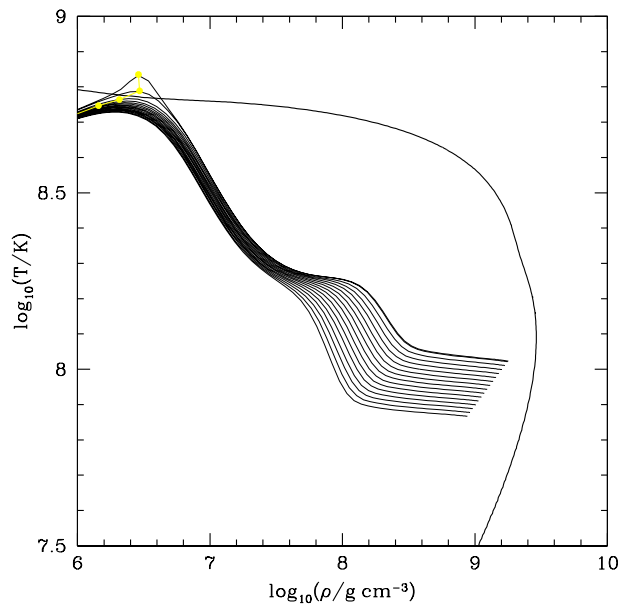}
\caption[]{As figure~\ref{ign} but beginning with the dashed line of
figure~\ref{initial}.  Ignition occurs close to the surface
and carbon ignites under much less degenerate conditions.  The
accretion rate is $3\times 10^{-6}\rm \, M_{\odot}yr^{-1}$. Grey
filled circles indicate the path of the mass shell where carbon first
ignites at mass $1.354\,M_\odot$ just $0.008\,M_\odot$ below the surface.}
\label{ign2}
\end{figure}

When carbon ignites in a thermonuclear runaway it manifests itself by
a sudden rise in carbon burning luminosity followed by breakdown of
the evolution code.  Figure~\ref{initial} shows how the internal run
of temperature and density from the cool surface to hot relatively
isothermal interior changes during accretion at rates just above and
just below the critical rate at which ignition moves from the centre
to the surface.  To be sure of what has happened, we determine a
carbon ignition curve as a function of temperature and density by
equating the carbon burning energy production with that lost in
neutrinos -- see Figures \ref{ign} and~\ref{ign2}.  Above and to the
right of this line we have the right conditions for carbon ignition to
run away.  Below it, any carbon burning ceases if the accretion is
turned off.  On the same axes in figures 2 and~3 we plot the internal temperature against
density for the models leading up to ignition.  If a line from our
internal model crosses the ignition curve then we know that the carbon
in the star has ignited at that particular point in the model.  There
is then a clear distinction between those models that ignite at the
centre (figure~\ref{ign}) and those which ignite in the outer layers just below the
surface (figure~\ref{ign2}).  There is not a smooth transition of the ignition point from
the centre to the surface because, in both cases, the intervening
parts of the white dwarf lie well away from the ignition curve.  The
transition is sudden and so the critical rate at which it occurs is
well determined for each of the cases we consider.  We note  further that this
ignition in the outer layers differs from the concept of an off-centre but still
relatively central ignition in the degenerate interior which could still
lead to thermonuclear runaway and supernovae.

\begin{table*}
\begin{tabular}{|c|c|c|c|c|c|c} \hline
ZAMS Mass/$M_{\odot}$  & WD Mass/$M_{\odot}$&Carbon &Oxygen &$\log_{10}( T_c/ \rm K)$ &   $\dot M_{\rm crit}/10^{-6}\rm M_{\odot}yr^{-1}$   \\\hline
5 & 0.50 & 0.21745 & 0.75821 & 7.0459 & 3.3 \\ 
5 & 0.50 & 0.21745 & 0.75821 & 7.3412 & 3.3 \\ \hline
5 & 0.70 & 0.21745 & 0.75821 & 6.9414 & 2.8 \\  
5 & 0.70 & 0.21745 & 0.75821 & 7.5093 & 3.1 \\ 
5 & 0.70 & 0.21745 & 0.75821 & 8.1160 & 3.1 \\ \hline
5 & 1.00 & 0.21744 & 0.75822 & 7.1625 & 2.7 \\ 
5 & 1.00 & 0.21744 & 0.75821 & 7.4911 & 3.0 \\ \hline
5 & 1.20 & 0.21744 & 0.75822 & 7.3499 & 2.8 \\
5 & 1.20 & 0.21744 & 0.75822 & 7.4937 & 2.7 \\ 
5 & 1.20 & 0.21744 & 0.75822 & 7.6290 & 2.8 \\
5 & 1.20 & 0.21744 & 0.75822 & 8.1199 & 3.1 \\ \hline
4 & 0.70 & 0.19659 & 0.77913 & 6.9480 & 2.8 \\
4 & 0.70 & 0.19659 & 0.77913 & 7.2597 & 3.1 \\ \hline
2 & 0.36 & 0.14199 & 0.83387 & 7.8905 & 3.2 \\ \hline
2 & 0.40 & 0.14199 & 0.83387 & 7.4963 & 3.2 \\ \hline
2 & 0.70 & 0.14199 & 0.83387 & 7.5106 & 3.2 \\ \hline
2 & 1.00 & 0.14197 & 0.83389 & 7.1647 & 2.8 \\
2 & 1.00 & 0.14197 & 0.83388 & 7.5003 & 3.1 \\
2 & 1.00 & 0.14199 & 0.83387 & 8.1182 & 3.2 \\ \hline
2 & 1.10 & 0.14196 & 0.83390 & 7.2220 & 2.8 \\ \hline
2 & 1.20 & 0.14193 & 0.83393 & 7.3406 & 2.9 \\
2 & 1.20 & 0.14193 & 0.83392 & 7.4975 & 2.8 \\
2 & 1.20 & 0.14194 & 0.83392 & 7.9003 & 3.2 \\
2 & 1.20 & 0.14195 & 0.83391 & 8.1181 & 3.2 \\ \hline
\end{tabular}
\caption{Critical constant accretion rates.  Column~1 is the original
zero-age main-sequence mass  
of the star from which the white dwarf was made.  Column~2 is the
white dwarf mass. It's central carbon and oxygen abundances are given
in columns~3 and~4.  The initial central temperature $T_{\rm c}$ is
in column~5. Column~6 is $\dot M_{\rm crit}$ the critical accretion
rate at which the ignition switches from the centre to the outside of
the white dwarf.
\label{tab1}}
\end{table*}

\section{Constant Accretion Rates}
\label{seccon}
\citet{nomoto1985} considered accretion at constant rates.  For a direct
comparison we first examine a range of initial conditions and constant
accretion rates (as in figures \ref{ign} and~ \ref{ign2} to estimate the critical rate at which ignition
switches from centre to surface and how it depends on initial
temperature and central carbon to oxygen ratio.  Table~\ref{tab1}
shows the critical accretion rates at which the ignition switches from
the centre to the outside of the star for different white dwarf
models.  We see that, in general, the hotter the initial central
temperature, the higher the critical accretion rate. The hotter
central temperature makes it easier for carbon to ignite at the centre
before the outside. For white dwarfs of similar temperature, the
higher the initial mass, the lower the critical accretion rate. Also,
the higher the carbon to oxygen ratio, the lower the critical
accretion rate.

\section{Eddington Accretion Rates}
\label{secedd}

Because it has some physical meaning and could place a real limit on
the accretion rate, it is also interesting to investigate the response
of a white dwarf to accretion that varies at a fixed fraction of the
Eddington rate $\dot M_{\rm EDD}$.
\cite{nomoto1985} found the critical rate to be about one fifth of $\dot
M_{\rm EDD}$. In the case of one dimensional spherical accretion the
kinetic energy of the accreting material is liberated in a shock at
the surface of the white dwarf and, if radiative transfer is the only
mechanism for the energy to escape, the consequent radiation pressure
slows the infalling material.  A maximum accretion rate, $\dot M_{\rm
EDD}$, is reached when the radiation force on particles balances
gravity at the surface of the white dwarf so
\begin{equation}
\dot M_{\rm EDD}=\frac{4\pi c}{\kappa}R_{1},
\end{equation}
where $R_1$ is the radius of the accreting white dwarf, $c$ is the
speed of light and $\kappa$ is the opacity of the accreting material
which is dominated by electron scattering.  This is found by equating the
Eddington luminosity and the accretion luminosity, that released by
material falling from infinity to the surface of the white dwarf.  The
Eddington limit decreases as $M_1$ increases because the radius $R_1$
falls as $M_1$ rises.
\par
In all cases accretion at $\dot M_{\rm EDD}$ is sufficiently fast to
cause ignition in the outer layers and so we also consider accretion
at a fraction of $\dot M_{\rm EDD}$.  In spherically symmetric accretion
$\dot M_{\rm EDD}$ represents the absolute limit to the rate at which
mass can be accreted.
\citep{han1999} make a more careful analysis, taking into account
the potential difference between the inner Lagrangian point and the
white dwarf surface and so permit higher accretion rates.  If
accretion is not spherically symmetric so that material can accrete
along some directions and radiation escape in others even higher
accretion rates could be possible.  On the other hand, if the opacity
$\kappa$ increases rapidly as material is driven off, accretion may be
limited to a lower rate in a similar manner to the limit on hydrogen
accretion in the model of \citet{hachisu1996}.

\begin{table*}
\begin{tabular}{|c|c|c|c|c|c|c} \hline
ZAMS Mass/$M_{\odot}$   & WD Mass/$M_{\odot}$&Carbon &Oxygen &$\log( T_c / \rm K)$ & Critical fraction   \\ \hline
5 & 1.000 & 0.21744 & 0.75822 & 7.1625 & 0.31 \\ 
2 & 1.100 & 0.14196 & 0.83390 & 7.2220 & 0.36 \\
2 & 1.200 & 0.14193 & 0.83393 & 7.3406 & 0.38 \\
5 & 1.300 & 0.21743 & 0.75823 & 7.5907 & 0.39 \\ \hline
\end{tabular}
\caption{Headings as for table~\ref{tab1} except the last column
which is the fraction of the Eddington accretion rate for which the
carbon first ignites in the outside of the white dwarf.\label{eddtab}}
\end{table*}

\begin{figure}
\epsfxsize=8.4cm
\epsfbox{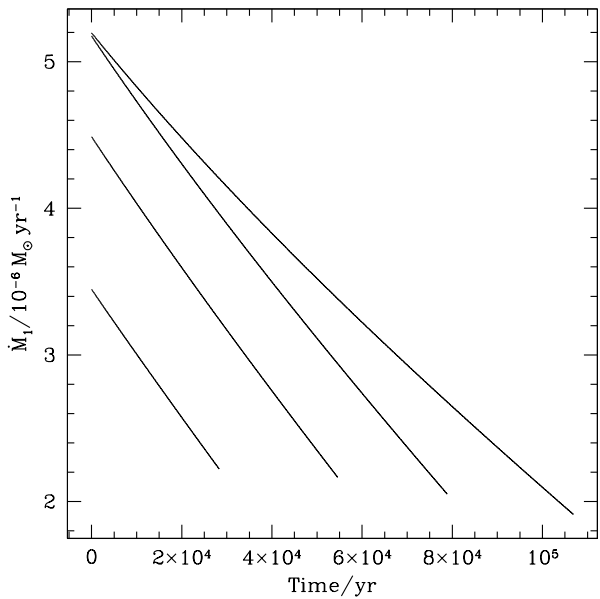}
\caption[]{The variation with time of accretion rate at the critical fractions
of the Eddington rate for the four cases listed in
table~\ref{eddtab}.  The rate is proportional to the radius of the
white dwarf and so falls off as mass increases.  The top line
is for accretion on to the white dwarf of initial mass $1\,M_\odot$
and the bottom of $1.3\,M_\odot$}
\label{eddrates}
\end{figure}

The radius $R_1$ depends on the mass $M_1$ of the accreting white
dwarf according to the formula of \citet{nauenberg1972} as
\begin{equation}
R_{\rm 1}= 0.0115\sqrt{ \left( \frac{ M_{\rm
ch}}{M_1}\right)^{2/3}-\left(\frac{M_1}{M_{\rm ch}}\right)^{2/3}} \,\rm
R_{\odot} \label{rad}
\end{equation}
where $M_{\rm ch}=1.44\,M_\odot$ is the Chandrasekhar mass.  For a
selection of white dwarfs table~\ref{eddtab} lists the critical
fraction of $\dot M_{\rm EDD}$ at which we move from central to
outside ignition and Figure~\ref{eddrates} shows how this critical
rate varies with time for each case.  We see that we can accrete up to
about two fifths of $\dot M_{\rm EDD}$ and still find
central ignition.  This is double the rate estimated by
\cite{nomoto1985}.  The higher the initial mass of the white dwarf the
higher the critical fraction of the Eddington rate.  We do not profess
to have identified an actual physical mechanism that would limit the
accretion in this way but rather wish to indicate what would be
necessary.

\section{Roche lobe overflow in white dwarf systems}
\label{secbin}

The actual rate of mass transfer between double white dwarfs often
exceeds $\dot M_{\rm EDD}$ but it is not constant.  Rather it falls
off as mass is transferred and the system widens in response to the
expansion of the radius of the mass-losing (donor) white dwarf.  The
separation is controlled by two competing processes.  Gravitational
radiation acts to shrink the orbit while mass transfer from the less
to more massive star acts to widen it.  For stable mass transfer the
Roche lobe must expand faster than the white dwarf, which grows in
response to mass loss.  The Roche lobe itself grows because the system
expands as mass in transferred to the more massive companion.  The
combination of these changes determines the mass-transfer rate
necessary to maintain $R_2 = R_{\rm L2}$ and the so the system widens.
As it widens the system experiences weaker gravitational-radiation
braking and the mass-transfer rate falls.

We set up a simple model that allows us to determine this
mass-transfer rate and limit it to a maximum fraction $f$ of $\dot
M_{\rm EDD}$.  In this way we can find systems in which the primary
white dwarf ignites at the centre even though the initial mass
transfer rate might be considerably higher than the critical rates
calculated earlier.

In a binary system the Roche lobe radius $R_{{\rm L}i}$ is the radius
of a sphere which encloses the same volume as that enclosed by the
last stable equipotential surface around star~$i$ ($i=1,2$) as given by
\citep{Egg83},
\begin{equation}
\frac{R_{{\rm L}i}}{a}=\frac{0.49q_i^{2/3}}{0.6q_i^{2/3}+\log(1+q_i^{1/3})} ,
\end{equation}
where $q_i$ is the mass ratio $M_i/M_{3-i}$ and $a$ is the separation
of the two stars. This formula is accurate to $1$ percent for all $q$.

In a binary system of two white dwarfs, when the less massive, of
mass $M_2$ and radius $R_2$, overfills its Roche lobe it loses
mass, some of which is accreted by its companion, of mass
$M_1$ and radius $R_1$.  For stable mass transfer the
rate ($-\dot M_2$) is determined by the angular momentum loss
from the orbit because $\dot M_2$ adjusts to keep $R_2\approx R_{\rm
L2}$ and $\dot R_2\approx \dot R_{\rm L2}$.
We model the mass-loss rate from the donor
white dwarf by
\begin{equation}
\dot{M_2}=- \dot M_0 e^{\frac{\Delta R}{H}}, \label{m2dot}
\end{equation}
where $\Delta R$ is the amount by which the white dwarf overfills its
Roche lobe,
\begin{equation}
\Delta R= R_2-R_{\rm L2}
\end{equation}
and $H$ is the pressure scale height \citep{ritter1988,martin2005}.
This formula takes account of the fact that stars have a thin
atmosphere above their photosphere and Roche lobe overflow begins
while the photosphere itself is still below the inner Lagrangian
point.  It
also allows us to make a smooth
transition to stable mass transfer as the white dwarfs are driven
together.  Because it is a steep function of $\Delta R$ we have $R_2
\approx R_{\rm L2}$ once mass transfer has begun.  The constant
$\dot M_0$ is the mass-transfer rate when the donor star fills its Roche lobe exactly so
that $\Delta R=0$.
We use $\dot M_0=3\times 10^{-6}$.
This is approximately the critical mass transfer rate where ignition
moves from the centre to the outside of the WD. If this rate is
exceeded then we slightly over estimate the mass transfer rate because
the stars are slightly closer than they would be if $\Delta R=0$ and
$|\dot J_{\rm gr}|$ (see equation~\ref{gr} below) is slightly larger.  If
$\dot M<\dot M_0$ we slightly underestimate the rate.  Figure~\ref{mt}
below shows that this rate is typical of the systems in which we are
interested.

We take $H$ to be a constant fraction of one thousandth of the radius
of the losing white dwarf.  Such a small fraction is typical of white
dwarfs so that $R_2$ is indeed very close to $R_{\rm L2}$.  Any
variation in $H$ is unimportant because we are only interested here in
the equilibrium mass-transfer rate.

We limit the mass accretion rate to $f\dot M_{\rm EDD}$,
where $0<f<\infty$ is a constant.  A factor $f \gg 1$ allows all the
mass transferred to be accreted ($\dot M_1 = -\dot M_2$) while $f =1$
applies the Eddington limit.
If the mass-transfer rate
is less than $f\dot M_{\rm EDD}$ all of the mass lost by the donor is accreted on
to the companion.  In general the accretion rate on to star $1$ is given by
\begin{equation}
\dot M_1 = \min (f\dot M_{\rm EDD},-\dot M_2 ). \label{m1dot}
\end{equation}
When the limit is active there is mass loss from the system because
star $2$ is losing mass more quickly than star $1$ is gaining it and
$\dot M=\dot M_1 + \dot M_2\le 0$.  This mass is blown away from the
binary in a wind which carries angular momentum with it.  We
assume it has the specific angular momentum of the accreting white
dwarf whence it is blown off and so the rate of loss of angular
momentum is
\begin{equation}
\dot J_{\rm wind} = \dot M \left( \frac{M_2}{M} a\right)^2 \Omega,
\end{equation}
where $\Omega$ is the orbital angular velocity.  For these close
systems loss of angular momentum owing to gravitational radiation is
the mechanism driving the evolution.  For two point masses in a
circular orbit the rate of change of angular momentum $\dot J_{\rm gr}$
is given by \cite{landau1951}
\begin{equation}
\label{gr}
\frac{\dot{J}_{\rmn{gr}}}{J}=-\frac{32G^3}{5c^5}\frac{M_1M_2(M_1+M_2)}{a^4}.
\end{equation}
We assume that there are no other angular momentum losses from the
system so that the total angular momentum loss is
\begin{equation}
\dot J_{\rm total}=\dot J_{\rm wind}+\dot J_{\rm gr}. \label{jdot}
\end{equation}

\begin{figure}
\epsfxsize=8.4cm
\epsfbox{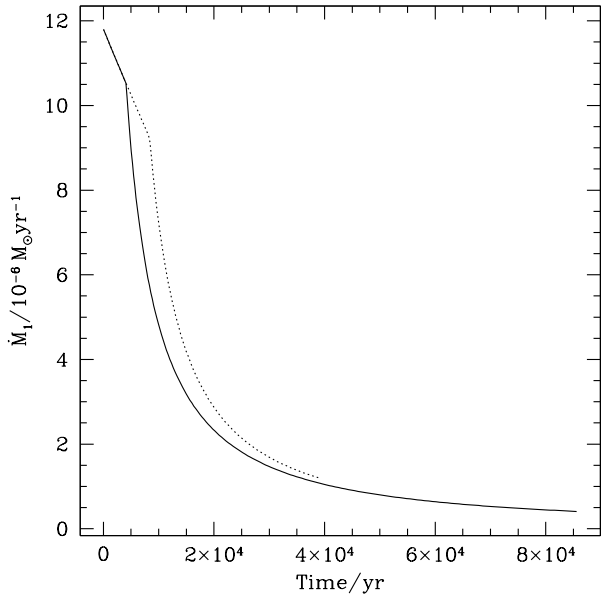}
\caption[]{Accretion rates on to a $1.2\,\rm M_{\odot}$
white dwarf.  The solid line is from a white dwarf companion of mass
$0.3\,\rm M_{\odot}$ and the dotted line of mass $0.9\,\rm
M_{\odot}$. Both are limited by $\dot M_{\rm EDD}$ and this limit,
which depends only on the mass of the gainer, is
effective over the initial straighter section. Both tracks end when the
accretor would have reached $1.38\rm M_{\odot}$.  Note that no account
of carbon burning has yet been taken.}
\label{mt}
\end{figure}

We can solve the three differential
equations~(\ref{m2dot}), (\ref{m1dot}) and~(\ref{jdot}) with a Runge
Kutta method and hence find the accretion rate.
Figure~(\ref{mt}) shows an example of accretion rates on to the
same mass of accreting white dwarf from two different mass companions.
The rates both begin at the Eddington rate and fall off in such
a way that the accretion rate from a lower mass companion is always
less than or equal to that from a higher and the time spent at rates
above the critical rates we reported in sections \ref{seccon}
and~\ref{secedd} is
significantly reduced for the lower mass companion.  They
both switch to accreting all of the mass that is lost from the
companion once this rate is less than the Eddington rate.

We note here that, had we not used an Eddington limited accretion
rate, the binary of $1.2\,\rm M_{\odot}$ and $0.9\,\rm M_{\odot}$
white dwarfs would have begun dynamically unstable mass transfer.  The
limit on accretion rate has a stabilizing effect on the mass transfer.

\section{Carbon ignition in binary star models}
\label{secsup}

We now consider detailed evolution of the accretion at variable rates
according to the binary model.  We begin with a separation such that
the lower-mass white dwarf is about to fill its Roche lobe.  The
initial white dwarfs are those described in section~\ref{seccon} and used in
section~\ref{secedd}.  In order for accretion to switch on smoothly by
equation~(\ref{m2dot}) we must prime our white dwarf model by rapidly
building up the accretion rate over several short time steps to about
$10^{-6}\,\rm M_\odot\, yr^{-1}$ as described in section~\ref{seccode}.  This means our surface layers are
slightly hotter than they should be at the onset of accretion but the
effect is soon swamped and could in any case only reduce the
likelihood of central ignition.

\begin{figure}
\epsfxsize=8.4cm
\epsfbox{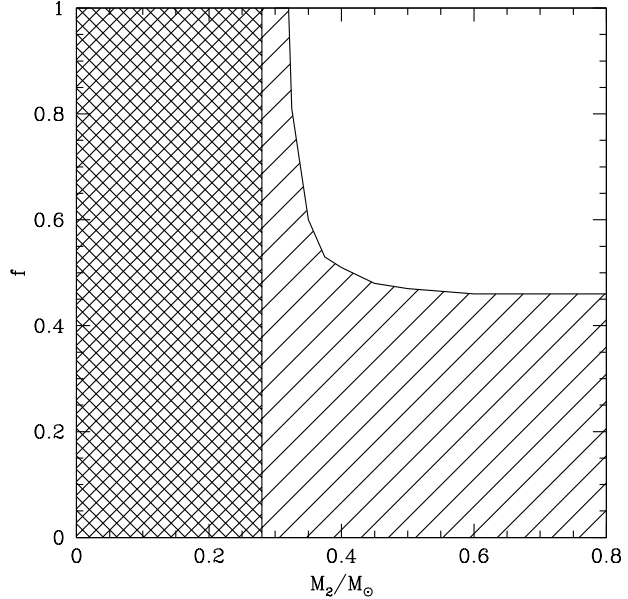}
\caption[]{Outcome for a $1.1\,\rm M_{\odot}$ white dwarf accreting
from a companion white dwarf of initial mass $M_2$ with an upper limit
to the accretion rate of $f\dot M_{\rm EDD}$.  In the single shaded
area the white dwarf ignites at the centre.  The blank area it
accretes too fast and ignite at the outside.  In the cross-hatched
area the initial total mass $M = M_1 + M_2 \le 1.38\,M_\odot$ so that
central ignition cannot be reached.}
\label{pic11}
\end{figure}

\begin{figure}
\epsfxsize=8.4cm
\epsfbox{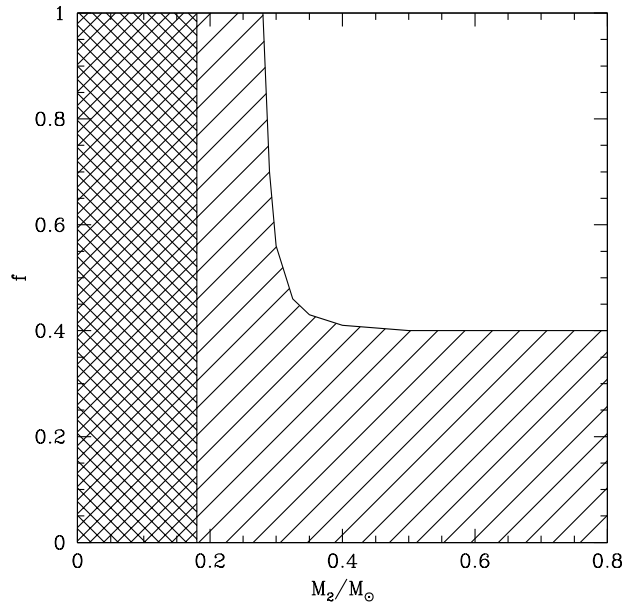}
\caption[]{Labels as for figure \ref{pic11} for a $1.2\,\rm M_{\odot}$
accreting white dwarf.}
\label{pic12}
\end{figure}

Figures~\ref{pic11} and~\ref{pic12} illustrate these, our most
realistic models, for initial mass accretors of $1.1$ and~$1.2\,\rm
M_\odot$ of the compositions given in table~\ref{eddtab}.  If the
total mass of the system is less than $1.38\,\rm M_\odot$ central
ignition can't occur so the cross-hatched region is always ruled out.
To the right of this the models in the single shaded regions show central
carbon ignition while those in the unshaded regions ignite at the
surface.  For a white dwarf of mass $1.1\,\rm M_{\odot}$ we find
central ignition for any mass companion with an accretion rate
$46\,$per cent of the Eddington limit. Similarly for a $1.2 \,\rm
M_{\odot}$ accreting white dwarf we can succeed with any mass
companion with accretion limited to $40\,$per cent of the Eddington
rate.  The curves rise steeply as $f$ approaches~$1$ and the mass
transfer can be fully conservative for lower-mass companions and still
lead to central ignition.

\begin{figure}
\epsfxsize=8.4cm
\epsfbox{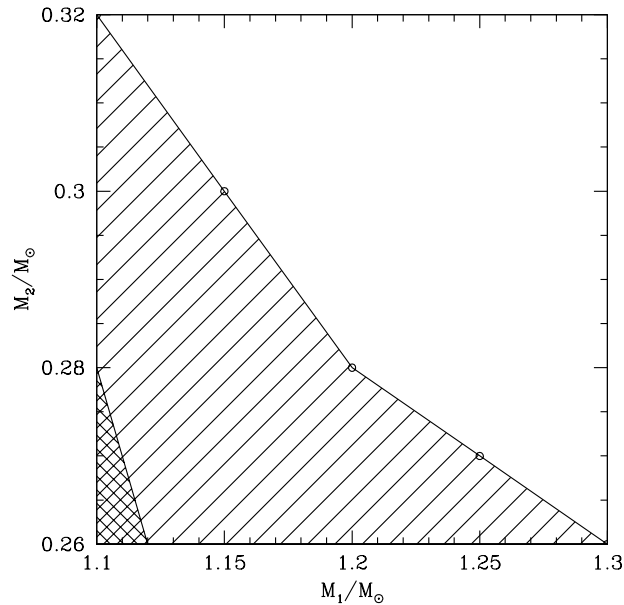}
\caption[]{The maximum mass of donor white dwarf $M_2$ for which we
find central ignition of a white dwarf of initial mass $M_1$ with
fully conservative mass transfer $f\gg 1$.  Circles are the masses
actually tested.  Systems in the single shaded region can reach central
ignition.  Those in the cross-hatched region do not have enough mass.
Note that the result is very similar if $f = 1$.}
\label{max}
\end{figure}

Figure~\ref{max} shows the highest mass of the donating white dwarf
for which we find central ignition for a given mass of the accreting white
dwarf with fully conservative evolution ($f\gg 1$).  Above this mass
the mass-transfer rate is too high and we find carbon ignition at the
outside and consequently no supernova.

\section{Conclusions}
\label{secconc}

Using detailed models of accreting CO white dwarfs we have critically
investigated what rates allow central degenerate carbon ignition as
opposed to ignition near the surface.  To compare with earlier work we
first considered accretion at constant rates and then at fractions of
the Eddington-limited rate.  Our results are qualitatively similar to
those of \cite{nomoto1985} but we find central ignition at rates up to twice
what they did or two-fifths of $\dot M_{\rm EDD}$.  This can be attributed
to their relatively course grid and updated stellar physics.
We have also considered real double white dwarf
binaries with mass transfer by Roche lobe overflow.
Variation in the initial
temperature and composition of the accreting white dwarf have only a
small effect.
We note that with
fully conservative mass transfer a $1.1\,M_\odot$ CO white dwarf with
a companion CO white dwarf of $0.3\,M_\odot$ can reach the conditions
required for a SN~Ia even though the accretion rate initially rises to
a maximum of almost $1.7\times 10^{-5}\, \rm M_{\odot}yr^{-1}$ and
carbon burning in the non-degenerate layers reaches a luminosity of
nearly $21\, \rm L_{\odot}$.  The mass transfer rate and hence the
luminosity fall off sufficiently quickly with the mass-transfer rate
that the burning is extinguished before it runs away and burns the
entire white dwarf.

Though standard single star evolution with reasonable mass loss only
produces CO white dwarfs from about $0.55\,M_{\odot}$ to $1.2\,M_{\odot}$
we note that a star of initial mass $2.5\,M_{\odot}$ has a CO core of
$0.28\,M_\odot$ at the start of the asymptotic giant branch,
immediately after convective core helium burning.  We also recall that
one or two phases of common envelope evolution are required to bring
the two white dwarfs close enough for gravitational radiation to
operate and that it would be easy to strip such a star's hydrogen rich
envelope to leave a naked helium star of about $0.55\,M_\odot$.  Such a
star is very small and would not undergo further interaction until
gravitational radiation sets in but it is also very luminous (about
$200\,L_\odot$) and maintains this luminosity for a long time (about
$5\times 10^8\,$yr owing to helium burning.  This could well drive off the helium rich material,
in a similar way to Wolf-Rayet stars, and expose the required low-mass
CO core.  \cite{hamann1995} suggest a mass-loss rate from naked helium
stars of as much as $10^{-11.95}(L/L_\odot)^{1.5}$ which is
considerably more than would be needed here.

Undoubtedly the parameter space leading to these conditions is small.
In the population synthesis of \cite{hurley2002} only one in five
hundred of the merging CO white dwarfs would actually make it to
central carbon ignition under these conditions and thence a supernova
rate of only about a thousandth of that observed.  However
much of the physics, including that of common envelope evolution and
mass loss from naked helium stars is highly uncertain, as are the
initial mass functions, mass ratio distributions and separation
distributions of binary stars so any population synthesis model is
still very uncertain.  We note also that interactions between stars in
dense stellar environments can significantly increase the number of
merging white dwarfs.  \cite{shara2002} find a factor of fifteen
increase in open clusters.

On the other hand if the mass accretion rate can be limited to $0.46$ of
$\dot M_{\rm Edd}$ then a $1.1\,M_\odot$ white dwarf accreting from a
companion of any mass above $0.28\,M_\odot$ can make it to a
supernova.  This would indeed be the case if a disc wind similar to
that postulated by \citet{hachisu1996} operated in the
double-degenerate case in a similar way to the single-degenerate
model.

We conclude that double CO white dwarfs remain as viable progenitors
of SNe~Ia as any others currently proposed but are unlikely to be
responsible for the majority of SNe~Ia.  In particular, if the
accretion luminosity were somehow limited to less than 46~per cent of
the Eddington rate in the early stages, then stable mass transfer
could account for an important fraction of observed Sne~Ia.

\section*{Acknowledgements}
We are grateful to John Lattanzio and Monash University for their
hospitality during which a substantial amount of this work was carried
out. RGM thanks Churchill College for travel grants to do this
work. CAT thanks Churchill College for a fellowship.  We thank the
referee for carefully reading the manuscript and pointing many
opportunities for improvement.

\label{lastpage}

\begin{thebibliography}{99}
\bibitem[\protect\citeauthoryear{Anders \& Grevesse}{1989}]{ander} Anders E., Grevesse N., 1989, Geochim. Cosmochim. Acta, 53, 197
\bibitem[\protect\citeauthoryear{Alexander \& Ferguson}{1994}]{alex} Alexander D. R., Ferguson J. W., 1994, ApJ, 437, 879
\bibitem[\protect\citeauthoryear{Baade}{1938}]{Baa} Baade W., 1938, ApJ, 88, 285
\bibitem[\protect\citeauthoryear{Branch}{1998}]{branch1998} Branch D., 1998, ARA\&A, 36, 17
\bibitem[\protect\citeauthoryear{Branch et al.}{1995}]{branch1995} Branch D., Livio M., Yungelson L. R., Boffi F. R., Baron E., 1995, PASP, 107, 717
\bibitem[\protect\citeauthoryear{Caughlan \& Fowler}{1988}]{cau} Caughlan G. R., Fowler W. A., 1988, At. Data Nucl. Data Tables, 40, 283
\bibitem[\protect\citeauthoryear{Eggleton}{1971}]{Egg71} Eggleton P. P., 1971, MNRAS, 151, 351
\bibitem[\protect\citeauthoryear{Eggleton}{1972}]{Egg72} Eggleton P. P., 1972, MNRAS, 156, 361
\bibitem[\protect\citeauthoryear{Eggleton}{1973}]{Egg73} Eggleton P. P., 1973, MNRAS, 163, 279
\bibitem[\protect\citeauthoryear{Eggleton}{1983}]{Egg83} Eggleton P. P., 1983, MNRAS, 268, 368
\bibitem[\protect\citeauthoryear{Hachisu, Kato \& Nomoto}{1996}]{hachisu1996} Hachisu I., Kato M., Nomoto K., 1996, ApJ, 470, L9
\bibitem[\protect\citeauthoryear{Hamann, Koesterke \& Wessolowski}{1995}]{hamann1995} Hamann W.-R., Koesterke L., Wessolowski U., 1995, A\&A, 229, 151
\bibitem[\protect\citeauthoryear{Han \& Webbink}{1999}]{han1999} Han Z., Webbink R. F., 1999, A\&A, 349, L17
\bibitem[\protect\citeauthoryear{Hurley, Tout \& Pols}{2002}]{hurley2002} Hurley J. R., Tout C. A., Pols O. R., 2002, MNRAS, 329, 897
\bibitem[\protect\citeauthoryear{Nomoto \& Iben}{1985}]{nomoto1985} Iben I., Nomoto K., 1985, ApJ, 297, 531
\bibitem[\protect\citeauthoryear{Iglesias, Rogers \& Wilson}{1992}]{Igles} Iglesias C. A., Rogers F. J., Wilson B. G., 1992, ApJ, 397, 717
\bibitem[\protect\citeauthoryear{Kahabka \& van den Heuvel}{1997}]{kahabka1997} Kahabka P., van den Heuvel E. P. J., 1997, ARA\&A, 35, 69
\bibitem[\protect\citeauthoryear{Kawai, Saio \& Nomoto}{1987}]{saio1987} Kawai Y., Saio H., Nomoto K., 1987, ApJ, 315, 229
\bibitem[\protect\citeauthoryear{Landau \& Lifshitz}{1951}]{landau1951} Landau L. D., Lifshitz E. M., 1951, The Classical Theory of Fields. Pergamon Press, Oxford
\bibitem[\protect\citeauthoryear{Martin \& Tout}{2005}]{martin2005} Martin R. G.,Tout C. A., 1983, MNRAS, 358, 1036
\bibitem[\protect\citeauthoryear{Nauenberg}{1972}]{nauenberg1972} Nauenberg M., 1972, ApJ, 175, 417
\bibitem[\protect\citeauthoryear{Nelemans et al.}{2000}]{nelemans2000} Nelemans G., Verbunt F., Yungelson L. R., Portegies Zwart S. F., 2000, A\&A, 360, 101 
\bibitem[\protect\citeauthoryear{Nelemans \& Tout}{2005}]{nelemans2005} Nelemans G., Tout C. A., 2005, MNRAS, 356, 753
\bibitem[\protect\citeauthoryear{Nomoto}{1982}]{nomoto1982} Nomoto K., 1982, ApJ, 253, 798
\bibitem[\protect\citeauthoryear{Paczy\'nski \& \.Zytkow}{1978}]{paczynski1978} Paczy\'nski B., \.Zytkow A. N., 1978, ApJ, 222, 604 
\bibitem[\protect\citeauthoryear{Piersanti et al.}{2003a}]{piersanti2003a} Piersanti L., Gagliardi G., Iben I., Tornamb\'e A., 2003, ApJ, 583, 885
\bibitem[\protect\citeauthoryear{Piersanti et al.}{2003b}]{piersanti2003b} Piersanti L., Gagliardi G., Iben I., Tornamb\'e A., 2003, ApJ, 598, 1229
\bibitem[\protect\citeauthoryear{Pols et al.}{1995}]{Pols} Pols O. R., Tout C. A., Eggleton P. P., Han Z., 1995, MNRAS, 274, 964
\bibitem[\protect\citeauthoryear{Ritter}{1988}]{ritter1988} Ritter H., 1988, A\&A, 202, 93
\bibitem[\protect\citeauthoryear{R\"opke \& Hillebrandt}{2005}]{ropke2005} R\"opke F. K., Hillebrandt W., 2005, A\&A, 431, 635
\bibitem[\protect\citeauthoryear{Shara \& Hurley}{2002}]{shara2002} Shara M. M., Hurley J. R., 2002, ApJ, 571, 830
\bibitem[\protect\citeauthoryear{Tout}{2005}]{tout2005} Tout C. A., 2005, in Hameury J.-M., Lasota J.-P. eds, ASP Conf. Ser. Vol.~330, The Astrophysics of Cataclysmic Variables and Related Objects. Astron. Soc. Pac., San Francisco, p.~279
\bibitem[\protect\citeauthoryear{Tout et al.}{2001}]{tout2001} Tout C. A., Reg\H os E.,Wickramasinghe D., Hurley J. R., Pols O. R., 2001, in Busso M., Gallino R eds, Salting the Early Soup: trace nuclei from stars to the solar system. Mem. Soc. Astron. Ital., 72, 371
\bibitem[\protect\citeauthoryear{Warner}{1995}]{warner1995} Warner B. 1995, {\it Cataclysmic Variables}, CUP, Cambridge
\bibitem[\protect\citeauthoryear{Webbink}{1984}]{Webbink1984} Webbink R. F., 1984, ApJ, 277, 355
\bibitem[\protect\citeauthoryear{Wickramasinghe \& Ferrario}{2005}]{wickramasinghe2005} Wickramasinghe D. T., Ferrario L., 2005, MNRAS, 356, 1576
\bibitem[\protect\citeauthoryear{Woolley \& Stibbs}{1953}]{Wool} Woolley R. v. d. R., Stibbs D. W. N., 1953, The Outer Layers of a Star (Oxford : Clarendon Press)
\bibitem[\protect\citeauthoryear{Woosley \& Weaver}{1994}]{woosley1994} Woosley S. E., Weaver T. A., 1994, ApJ, 423, 371

\end{thebibliography}
\end{document}